\documentclass[a4paper]{article}
\usepackage[font=scriptsize,labelfont=bf]{caption}
\usepackage{graphicx} 
\graphicspath{{figures/}} 
\usepackage{amsmath} 
\usepackage{amssymb}
\usepackage{hyperref}  
\usepackage[capitalise,nameinlink]{cleveref}
\usepackage{xcolor} 
\usepackage{tabularx}
\usepackage{gensymb}
\usepackage{booktabs}
\usepackage{microtype}
\usepackage{siunitx}
\usepackage{authblk}
\usepackage{pdfpages}
\title{Technical overview and architecture of the FastNet Machine Learning weather prediction model, version 1.0}
\date{September 2025}

\author[$\ast$1+]{Eric G. Daub}
\author[$\ast$2]{Tom Dunstan}
\author[$\ast$2]{Thusal Bennett}
\author[$\ast$2]{Matthew Burnand}
\author[$\ast$2]{James Chappell}
\author[$\ast$1]{Alejandro Coca-Castro}
\author[$\ast$1]{Noushin Eftekhari}
\author[$\ast$1,3]{J. Scott Hosking}
\author[$\ast$1]{Manvendra Janmaijaya}
\author[$\ast$2]{Jon Lillis}
\author[$\ast$1]{David Salvador-Jasin}
\author[$\ast$1]{Nathan Simpson}
\author[$\ast$1]{Oliver T Strickson}
\author[$\ast$1]{Ryan Sze-Yin Chan}
\author[$\ast$4]{Mohamad Elmasri}
\author[$\ast$1]{Lydia Allegranza France}
\author[$\ast$2]{Sam Madge}
\author[$\ast$2]{Aled Owen}
\author[1]{James Robinson}
\author[5,6]{Adam A. Scaife}
\author[2]{David Walters}
\author[1]{Peter Yatsyshin}
\author[2]{Theo McCaie}
\author[1]{Levan Bokeria}
\author[2,7]{Hannah Brown}
\author[2,8]{Tom Dodds}
\author[1]{David Llewellyn-Jones}
\author[2]{Sophia Moreton}
\author[2]{Tom Potter}
\author[1]{Iain Stenson}
\author[1]{Louisa van Zeeland}
\author[2,9]{Karina Bett-Williams}
\author[2]{Kirstine Ida Dale}

\affil[1]{The Alan Turing Institute, London, UK}
\affil[2]{Met Office, Exeter, UK}
\affil[3]{British Antarctic Survey, Cambridge, UK}
\affil[4]{Carleton University, Canada}
\affil[5]{Met Office Hadley Centre, Exeter, UK}
\affil[6]{Department of Mathematics and Statistics, University of Exeter, Exeter, UK}
\affil[7]{UK Hydrographic Office, Taunton, UK}
\affil[8]{Office for National Statistics, UK}
\affil[9]{Global Systems Institute, University of Exeter, Exeter, UK}
\affil[$\ast$]{Equal Contribution}
\affil[+]{Corresponding Author, edaub@turing.ac.uk}

\begin{document}
\includepdf{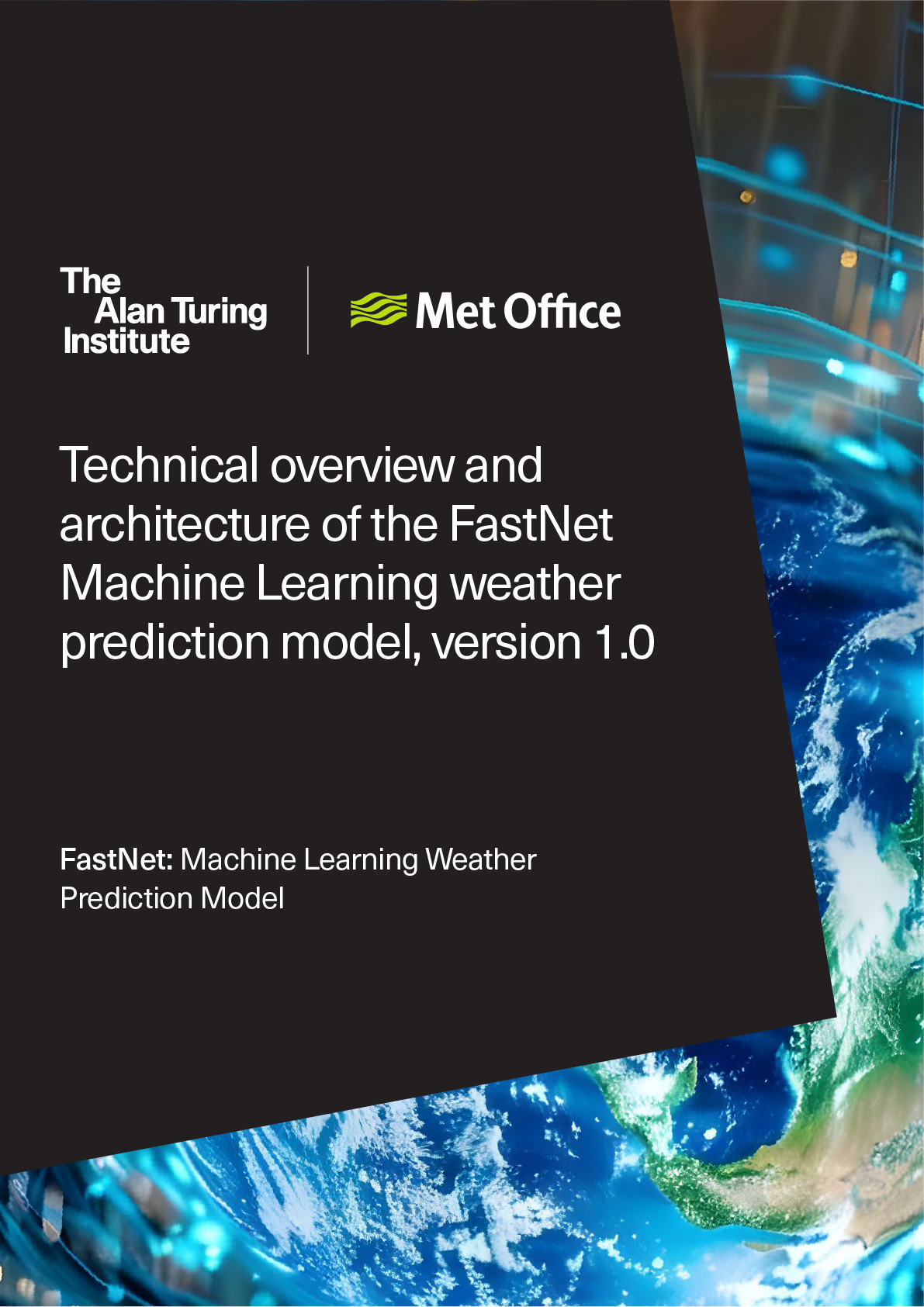}
\maketitle

\newpage

\abstract{We present FastNet version 1.0, a data-driven medium range numerical weather prediction (NWP) model based on a Graph Neural Network architecture, developed jointly between the Alan Turing Institute and the Met Office. FastNet uses an encode-process-decode structure to produce deterministic global weather predictions out to 10 days. The architecture is independent of spatial resolution and we have trained models at 1$\degree$ and 0.25$\degree$ resolution, with a six hour time step. FastNet uses a multi-level mesh in the processor, which is able to capture both short-range and long-range patterns in the spatial structure of the atmosphere. The model is pre-trained on ECMWF's ERA5 reanalysis data and then fine-tuned on additional autoregressive rollout steps, which improves accuracy over longer time horizons. We evaluate the model performance at 1.5$\degree$ resolution using 2022 as a hold-out year and compare with the Met Office Global Model, finding that FastNet surpasses the skill of the current Met Office Global Model NWP system using a variety of evaluation metrics on a number of atmospheric variables. Our results show that both our 1$\degree$ and 0.25$\degree$ FastNet models outperform the current Global Model and produce results with predictive skill approaching those of other data-driven models trained on 0.25$\degree$ ERA5.}

\section{Introduction}

Recent advances in data-driven medium-range weather prediction have shown that a variety of machine learning architectures produce global weather forecasts that match or surpass existing physics-based methods \cite{Pathak2022,pangu,Lam2023,keisler2022,Lang2024,arches2024}. Once trained on historical weather data, data-driven approaches can produce a forecast using a small fraction of the time and compute required to run the physics-based systems, with a level of accuracy that in many instances surpasses the skill of current state-of-the-art physics-based systems. To date, most of these models are deterministic \cite{Pathak2022,pangu,Lam2023,keisler2022,Lang2024,arches2024}, although ensemble methods based on machine learning are also being developed \cite{gencast,aifscrps}. While ensembles are generally required to fully capture the uncertainty around the range of expected weather patterns \cite{lewis-ensembles}, ensemble NWP approaches are often built by making perturbations to the initial conditions of a deterministic model \cite{ecmwf-ensembles}. Following on from the recent advances in data-driven medium-range weather prediction, the Met Office and the Alan Turing Institute have developed the FastNet data-driven NWP model and are working towards its implementation as part of an operational model used within the larger Met Office forecast systems. This report describes the technical design of the deterministic FastNet global model version 1.0 and demonstrates its predictive skill in comparison with the existing Met Office Global Model (GM) over a range of atmospheric variables and lead times.

FastNet uses a Graph Neural Network (GNN) architecture to produce a deterministic prediction of the future weather state and is trained on the European Centre for Medium Range Weather Forecasting (ECMWF)'s ERA5 reanalysis data from 1980--2021, using 2022 as a held-out evaluation year. We describe here two versions of FastNet: one trained on the $1\degree$ O96 reduced Gaussian grid and the other on the native ERA5 resolution $0.25\degree$ N320 reduced Gaussian grid. We have found that the $1\degree$ model currently produces our best results (when evaluated on a 1.5$\degree$ latitude-longitude grid), while only requiring a fraction of the compute required to train on the full native ERA5 resolution. Our results show that FastNet is able to produce skillful predictions over a range of variables and lead times when compared with the GM. The predictive skill of FastNet is also competitive with ECMWF's Integrated Forecast System-HRES physics-based model, and is comparable to other data-driven models trained on $0.25\degree$ ERA5.

The report is structured as follows. First, we provide an overview of the FastNet model and the technical decisions made in specifying and training the model. We then discuss how FastNet was evaluated using a set of deterministic metrics on the hold out year of 2022 for comparison with the Met Office Global Model. Finally we conclude with a brief discussion of future improvements and experimental operational testing of FastNet to continue to monitor its performance as future improvements and upgrades are made.

\section{Data}

\subsection{ERA5}

FastNet is trained and evaluated on the ERA5 reanalysis data set provided by ECMWF \cite{era5-paper}, which provides a comprehensive picture of historical weather on a uniform global spatial grid. ERA5 is produced by assimilating weather observations with ECMWF's Integrated Forecast System (IFS) physical model using the 4D-Var method \cite{rabier-4dvar}. Reanalysis methods use observations received both during and after the assimilation time window to constrain the state of the atmosphere at a given time and thus provides a comprehensive estimate of historical weather using the full set of available observations. ERA5 also uses a fixed model specification for the full dataset, rather than operational analyses which are based on a model that undergoes regular updates and improvements over time. FastNet uses data on 13 pressure levels taken at 6 hour time snapshots from the ERA5 archive. We use ERA5 computed on the native N320 (approximate grid spacing of 31 km) and re-gridded to an O96 (approximate grid spacing of 104 km) reduced Gaussian grids. Reduced Gaussian grids decrease the number of grid points at a given latitude with increasing latitude, aiming to maintain a consistent grid spacing over the entire globe when compared to regular latitude-longitude grids. This modification translates into a considerable reduction of the total number of grid points (and computation) compared to other weather data-driven models operating on regular longitude-latitude grids of similar resolutions. For instance, the N320 reduced Gaussian grid has \num{542080} grid points, while a $0.25\degree$ longitude-latitude grid has \num{1038240} grid points. We train versions of FastNet on both resolutions of ERA5 using the same model architecture, with differences noted as appropriate in the following sections of the text.

\subsection{Variables}

FastNet has three types of variables: forcing, surface, and level variables. The forcing variables are those that can be computed ahead of time and include land-sea mask, orography, top of atmosphere solar radiation, time, and latitude/longitude variables. Surface and level variables serve as both input and output of the model and serve as the main targets for training. There are no diagnostic (output-only) variables.

The full set of variables used in FastNet are shown in Table~\ref{tab:fastnet_variables}. This includes forcing, atmospheric variables defined on the full set of pressure levels, and near surface variables. Atmospheric variables are defined at the indicated pressure levels. This gives a full set of $5\times13 + 7 + 13=85$ total variables at each grid point. 
 
\begin{table*}[t]
\footnotesize
\centering
\caption{Variables used in FastNet O96 and N320 training.}
\begin{tabularx}{\textwidth}{p{0.48\textwidth}Xl}
\toprule
\textbf{Variable} & \textbf{Level} & \textbf{Input/Output} \\
\midrule
Geopotential\newline
Horizontal wind, zonal and meridional components\newline
Specific humidity \newline Temperature &Pressure levels (hPa):\newline 50, 100, 150, 200, 250, 300, 400, 500, 600, 700, 850, 925, 1000 &Both\\
\midrule
Surface pressure \newline Mean sea-level pressure \newline Skin temperature \newline 2m temperature \newline 2m dewpoint temperature \newline  10m horizontal wind, zonal and meridional components &Surface &Both\\
\midrule
Land-sea mask\newline Orography \newline Standard deviation of sub-grid orography \newline Slope of sub-scale orography \newline Top-of-atmosphere solar radiation \newline Solar hour angle (cos and sin) \newline Time of year (cos and sin) \newline Latitude (cos and sin) \newline Longitude (cos and sin) &Surface &Input-only \\
\bottomrule
\end{tabularx}
\label{tab:fastnet_variables}
\end{table*}

\subsection{Data Pre-processing and Normalisation}

Training data from each pressure level (including surface-level variables) are separately standardised to have zero mean and unit standard deviation, over latitude, longitude and time. Orography is rescaled to the unit interval, and land-sea mask, solar hour angle, time of year, latitude, and longitude require no change  as they are already suitably normalised.

\section{FastNet Model Architecture}

FastNet is a deterministic machine-learning weather model built in a purely data-driven way. The FastNet model is based on existing Graph Neural Network (GNN) architectures, most prominently GraphCast \cite{Lam2023} but also the work of Keisler \cite{keisler2022} and AIFS \cite{Lang2024}. FastNet comprises a series of GNNs arranged in an encode $\rightarrow$ process $\rightarrow$ decode structure. We include a short description of the mathematical operations underlying GNNs in appendix \ref{appendix:gnns}. The encoder takes data of the current atmospheric state, which we refer to as the \emph{grid}, and using a GNN maps the current atmospheric state to a lower-resolution latent space processor, which we refer to as the \emph{mesh}. The mesh has icosahedral symmetry, where we start with a base icosahedron (whose vertices are mapped to locations on the surface of the globe), and sub-divide each triangular face recursively until we reach the desired mesh resolution. The mesh state is advanced in time by six hour increments by the processor GNN, and the updated state is mapped back to the grid using a decoder GNN. To produce forecasts beyond six hours, the output grid state is autoregressively fed back into the model until the desired forecast lead time is reached.

FastNet uses a residual formulation, where the decoder output represents the increment (or residual) to be added to the input state, rather than predicting the full field from scratch. The predicted residuals are thus added to the original fields via a skip-level connection to produce the next weather state. This approach has been shown to perform better than directly predicting the field values at our step-size (6~h) \cite{Siddiqui2024}. Only a single weather state is used as input, as including additional forecast history did not improve model performance but does have an associated computational cost.

\subsection{Model Pipeline}

\begin{figure}[htbp]
  \centering
  \includegraphics[
      width=0.9\linewidth,
      keepaspectratio,
      trim={1cm 0.4cm 1cm 0.4cm},  
      clip
  ]{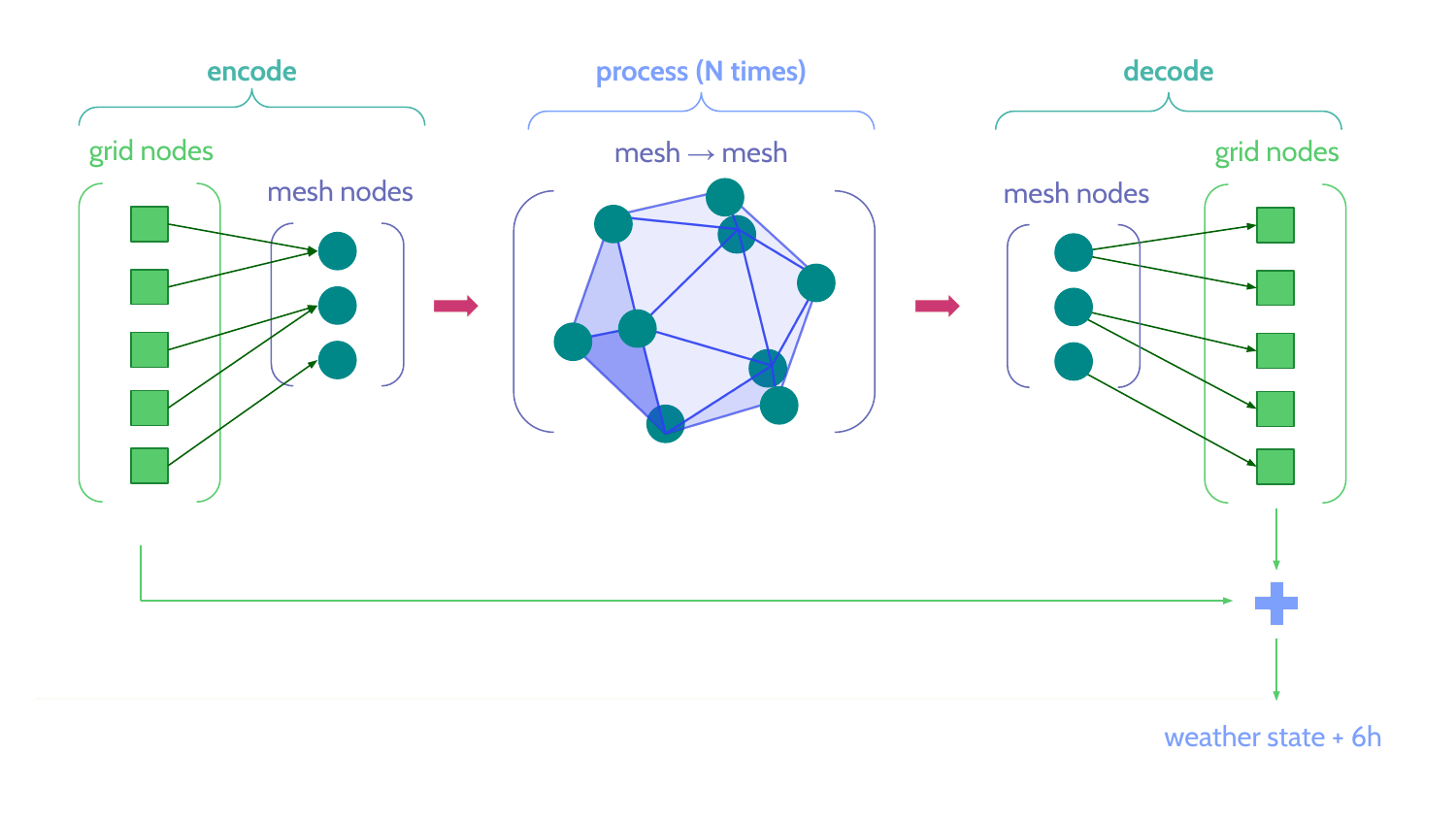}
  \caption{A high-level overview of the FastNet architecture. Data on the earth grid is first mapped to the mesh through an encode step via a Graph Neural Network, which embeds the grid features into a latent space. Once the weather state at $t=0$ is encoded onto the mesh, $N$ rounds of message-passing are carried out on the mesh to update the node and edge states via multi-level perceptrons (MLPs), with a separate MLP for each round. The state of the mesh is then decoded back to the grid, and added to the initial grid state as a residual connection. This advances the model forward one timestep ($t=6$~h). Longer forecasts are obtained by auto-regressively passing the model predictions back into the model and rolling out the forecasts over the desired lead time. We examine lead times up to 10 days, though our main comparisons with the UK Met Office GM are done out to a 7 day lead time. Note that the encoder and decoder graph connectivity in the figure are for illustrative purposes and are not representative of the actual model details; a full description of how these graphs are constructed is provided in the main text.}
  \label{fig:model_diagram}
\end{figure}

Here, we describe how the ERA5 gridded data is carried through a series of GNNs that encode, process, and decode the inputs. A high-level overview of the architecture is presented in \cref{fig:model_diagram}. Notably, this workflow involves two distinct, but coupled, encodings:

\begin{itemize}
    \item Mapping data inputs to a latent space
    \item Performing processing on a simple graph structure
\end{itemize}

These ideas were introduced in the context of the simulation of mesh dynamics in \cite{Pfaff2020}, and were adapted by \cite{keisler2022} for application to medium-range global weather prediction. The former gives the opportunity to learn useful representations for prediction, while the latter ensures the most computationally significant steps in the model are done in a lower resolution latent space on a mesh exhibiting uniform spatial coverage.

\subsubsection{Mesh}
Like other GNN models, we map our input data to an intermediate graph representing the state of the atmosphere, known as the mesh. This mesh consists of a set of overlapping icospheres. We generate the refined set of meshes using the \texttt{icosphere} package \cite{icosphere}, and parametrise those meshes through the desired angular resolution $\theta$.

\begin{figure}
\includegraphics[width=\textwidth]{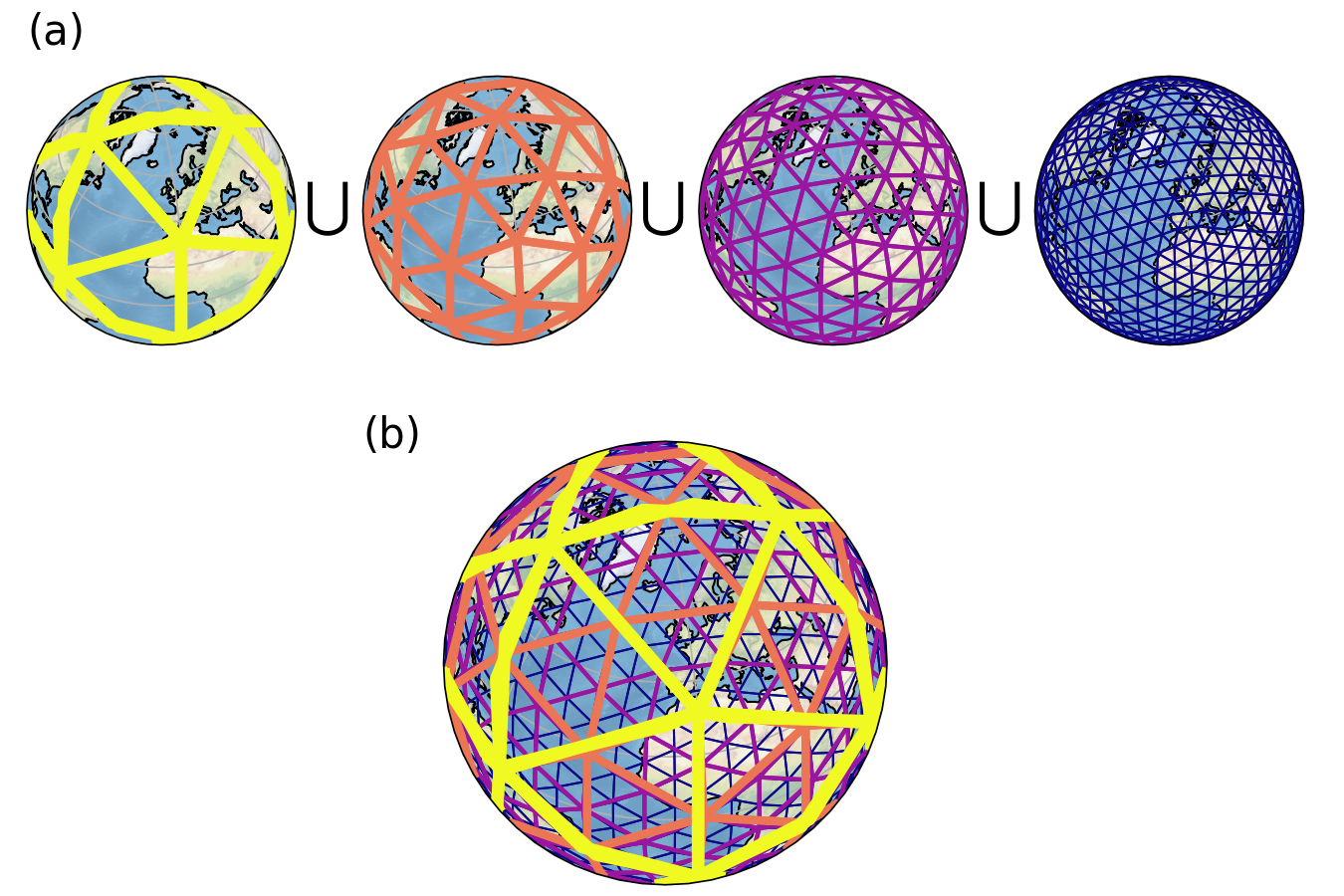}
\caption{Illustration of the multimesh used in FastNet \cite{Lam2023}. (a) Refined icosahedra used to construct the multimesh, including the base icosahedron and 3 levels of refinement. The $\approx 2\degree$ O96 mesh uses two additional refinement levels beyond the finest mesh shown, and the $\approx 1\degree$ N320 mesh uses three additional refinement levels. (b) Combined multimesh, resulting from unifying all mesh levels into a single structure using the finest mesh nodes but combining the node connections from all refinement levels. The combined multimesh includes both short-range and long-range connections that efficiently pass information across the globe.}
\label{fig:multimesh}
\end{figure}

Additionally, FastNet uses a multiscale mesh structure \cite{Lam2023}, illustrated in Fig.~\ref{fig:multimesh}. The multiscale mesh architecture is a structure that models Earth’s atmosphere with high spatial accuracy while still maintaining long-distance connectivity to efficiently pass information around the globe. Built on a refined icosahedron which has near-uniform node spacing, this mesh allows FastNet to avoid distortions inherent in latitude-longitude grids. Starting from a twelve node icosahedral graph, the mesh undergoes six levels of refinement, reaching $\num{40962}$ nodes at the finest scale, with each level reducing the node spacing by a factor of two.  The resolution of the resulting mesh is $\theta\approx1.26\degree$ for the model trained on N320 ERA5 native grid (six levels of refinement), and $\theta\approx2.52\degree$ for the O96 model (five levels of refinement).  Each level contains graph connections between all neighbouring nodes. 

The architecture integrates edges from all previous refinement levels in the unified multimesh, creating both short and long edges. Short edges connect nearby nodes for local detail, while long edges span greater distances, enabling the model to capture both localised and long-range interactions in one framework.

\subsubsection{Encoder}

The encoder is a directional bipartite graph from grid points to mesh points. The edge connectivity is determined by spatial proximity using one of two methods, which we compare.

The way in which information is transferred from grid to mesh is through treating this graph as a graph neural network, and performing one round of message passing. Since the edges are directed towards the mesh, only the mesh nodes will be updated with new messages. In particular, this is also where we define the size of our latent space; the MLP used to update the mesh nodes maps directly from the input data size to the latent space size of 768.  For comparison, this differs from GraphCast, where all variables on nodes and edges of the graphs involved are first embedded with a series of MLPs, and the encoder leaves that latent size unchanged when going from grid to mesh. We see equivalent performance for both methods, and thus default to the simpler direct MLP encoding.

The encoder aggregates information from the input grid onto the icosahedral mesh. Following architectures like AIFS \cite{Lang2024} and GraphCast \cite{Lam2023}, we construct the encoder graph by defining edges based on geographic proximity. Specifically, we consider two approaches for connecting the grid and mesh nodes: radius-based methods, where edges link each mesh node to all grid points within a fixed geodesic distance, and $k$-nearest-neighbour (KNN) methods, where each grid point is connected to the $k=2$ closest mesh nodes. Our FastNet model as presented here uses KNN to construct the graph, which we find gives the best performance for the model trained on O96 data. However, experiments using a radius-based method as used in AIFS and GraphCast gives slightly better performance on root mean-squared-error (RMSE) scores for the N320 model, defined in Eq.~\ref{eq:rmse}. For most output variables the RMSE improvement for the N320 model when using a radius-based encoder connectivity is approximately 2\% relative to the KNN model for long lead times. At shorter lead times, the RMSE differences between the models are similar in terms of absolute values, but are relatively larger when compared to the smaller RMSE values at short lead times, ranging from 5-10\% relative to the KNN RMSE value. More details on the differences between the KNN and radius graph connectivity can be found in Appendix~\ref{appendix:encoder}. We find that the choice of the encoder grid connectivity thus has only a small impact on the final model performance in terms of global RMSE scores; however there may be local artifacts due to the grid-mesh connectivity that affect the spatial patterns of model predictions that are more significant. Examining the spatial patterns of the mesh-grid connections in detail are beyond the scope of this report.

\subsubsection{Processor}

Once the current atmospheric state is mapped to the processor mesh, a GNN is used to advance the state forward with a six hour time step. The processor GNN is based on a the multi-mesh illustrated in Fig.~
\ref{fig:multimesh}, where the processor GNN combines edges from multiple lower resolution icosahedra, ensuring both short-range and long-range connections are able to capture the desired weather phenomena \cite{Lam2023}. The processor GNN contains 768 latent features at each mesh point, and performs 16 rounds of message passing when making its updates, with each round having its own set of learned parameters.

\subsubsection{Decoder}
The decoder uses a $k$-nearest-neighbour graph structure, where each grid point connects to the three nearest mesh nodes, a structure shared by GraphCast and AIFS. To map the mesh state back to the grid, the latent state at each mesh node is passed through an interaction network, and the output of the three nearest neighbours are concatenated for each grid point. An MLP then maps the resulting vector to the full set of atmospheric and surface variables at each grid point.

Since the whole formulation is trained end-to-end, all of the encoder, processing, and decoder networks are ``aware'' of each other through the passing of gradient information in backpropagation. This allows learned representations to efficiently encode information that is directly useful for predicting accurate forecasts, while simultaneously guiding the output to match ERA5 in the original gridded input space.

\subsection{Related models}

\label{sec:othermodels}

\textbf{GraphCast} and \textbf{FastNet} both employ 16-layer GNN processors that pass messages on multi-scale icosahedral meshes.  They are first trained on a single six hour time step and then fine-tuned autoregressively out to multiple lead times, following a curriculum learning approach to mitigate error compounding over longer lead times~\cite{Lam2023,Siddiqui2024}. 

\textbf{AIFS v1} replaces the earlier graph encoder (fixed-radius neighbours with distance–direction edge features) of AIFS v0.2 with a graph transformer that operates directly on an O96 Gaussian grid ($\sim\num{41000}$ points). Shifted-window attention within latitude bands and periodic window shifts emulate the locality–to–global mixing of a GNN while avoiding explicit edges; their forecast skill is similar to that of GraphCast~\cite{Lang2024}.  

\textbf{FourCastNet} performs convolution–Fourier mixing on a $0.25\degree$ lat–lon grid, leveraging global spectral coupling without an explicit encoder~\cite{Pathak2022}.  

\textbf{Pangu-Weather} applies a pure vision-transformer stack to 3-D tensors (latitude, longitude, levels), attaining performative forecast skill relative to IFS, but remaining tied to its native resolution. Pangu trains separate models over 1 hour, 3 hour, 6 hour, and 24 hour time horizons, which can be combined to forecast arbitrary lead times. This aims to reduce accumulation of forecast errors in long rollouts by reducing the model evaluations compared to other models that use a fixed time step \cite{pangu}.

At coarser resolutions, similar models include the GNN developed by \textbf{Keisler} \cite{keisler2022} at $1\degree$ and the transformer-based \textbf{ArchesWeather} \cite{arches2024} at $1.5\degree$, both of which have accuracy near or exceeding ECMWF's IFS-HRES but at a lower training cost when compared to the $0.25\degree$ models described above. Additionally, the \textbf{MET Norway/ECMWF stretched-grid GNN} varies node density from $\qty{2.5}{km}$ over the Nordics to $\qty{31}{km}$ elsewhere, so that a single graph captures both local high-resolution physics and global flow using a 16-layer core in line with other GNN models~\cite{Nipen2024}. 

\section{Training}

\subsection{Pre-Training}

The training objective in FastNet is a mean squared-error (MSE) between the target output $X$ and predicted output $\hat{X}$ for all output variables indicated in Table~\ref{tab:fastnet_variables} that combines errors by predicted variable and forecast lead time.

\begin{equation}
	\label{eq:mseloss}
\mathcal{L}_{MSE}=\frac{1}{\left|D_{\mathrm{batch}}\right|}\sum_{d_0\in D_{\mathrm{batch}}}{\frac{1}{T_{\mathrm{train}}}\sum_{\tau=1}^{T_{\mathrm{train}}}\sum_{j\in J}{s_jw_j}}\left({\hat{x}}_j^{d_0+\tau}\ -\ x_j^{d_0+\tau}\ \right)^{2\ }
\end{equation}
where, 
\begin{itemize}
    \item $\tau \in 1 : T_\mathrm{train}$ are the lead times being forecast by the model.
    \item $d_0 \in D_\mathrm{batch}$ represents forecast initialisation states within a given batch.
    \item $j \in J$ represents the variable index.
    \item $s_j$ represent the per-variable-level inverse variance of time differences.
    \item $w_j$ represent the per-variable-level pressure or surface weight.
    \item ${\hat{x}}_j^{d_0+\tau}$ and $x_j^{d_0+\tau}$ are the forecast and target values for some variable-level and lead time.
\end{itemize}

The contribution from each forecast field to the overall loss is first multiplied by a fixed weighting. These include, for forecast variables defined on pressure levels, a per-variable-level weighting proportional to the pressure level, 
\begin{equation}
    w_j = \frac{P_j}{\sum_j P_j},
\end{equation}
and a fixed value for surface variables of \num{1.0} for \qty{2}{m} temperature and \num{0.1} for every other field.

We also multiply by a weight corresponding to the per-variable-level inverse variance of time differences, 
\begin{equation}
    s_j = \mathbb{V}_{i,t} \left[x^{t+1}_{i,j} - x^t_{i,j} \right]^{-1},
\end{equation}
such that each variable, indexed by $j$, has a corresponding per-variable-level weight, $w_j \times s_j$. Because the reduced Gaussian grids have a reasonably uniform grid spacing over the entire globe, we do not area-weight the grid points in the loss calculation for the O96 or N320 models.

The O96 model is pre-trained on ERA5 for a single 6~h time step (i.e.~$T_\mathrm{train}=1$) using the AdamW optimiser with a peak learning rate of $10^{-3}$ for 100 epochs following a cosine schedule. We use a data parallel approach to split model training across 8 A100 GPUs with a local batch size of 2 per GPU, resulting in an effective batch size of 16 across the full model. Data parallel training involves making a copy of the model on each GPU, and splitting the data batches across the GPUs such that each GPU only sees a subset of the data. Each copy of the model computes the loss and its gradient independently for this subset of the data, and the loss and gradients are then combined to update the model parameters for each optimizer step. The O96 model requires 2.25 days to complete pre-training using this approach. The N320 model is pre-trained on ERA5 as above, with a peak learning rate of $2.5\times10^{-4}$ following a cosine schedule. The N320 model is pre-trained in a data parallel fashion on 24 GPUs with a local batch size of 1 per GPU, thus an effective batch size of 24 across the full model. The N320 model requires approximately 5.75 days to pre-train. 

\subsection{Fine-Tuning}

Following pre-training, we fine tune the models in a manner similar to GraphCast \cite{Lam2023} over longer forecast lead times from an autoregressive rollout. For the O96 model, we successively increase the number of autoregressive steps used in the loss function $T_\mathrm{train}$ from two to twelve steps while maintaining a small learning rate of $10^{-7}$. Each step in the O96 fine-tuning is trained for one epoch using a data parallel strategy, requiring 76 hours in total to complete the full set of fine-tuning stages. The N320 model is fine-tuned out to 7 autogregressive steps, with a slightly higher learning rate of $10^{-6}$. The N320 model is parallelised across 24 GPUs using a fully sharded data parallel strategy, and requires 47 total hours in total to complete the fine-tuning. Model sharding involves splitting the model parameters across multiple GPUs; this reduces GPU memory usage and enables computing the multi-step loss in Eq.~\ref{eq:mseloss} and its gradient. While model sharding reduces GPU memory usage on a single GPU, this comes at the expense of requiring additional all-to-all communication in both the forward and backward model pass steps in order to compute the loss and its gradient.

\begin{figure}
\includegraphics[width=\textwidth]{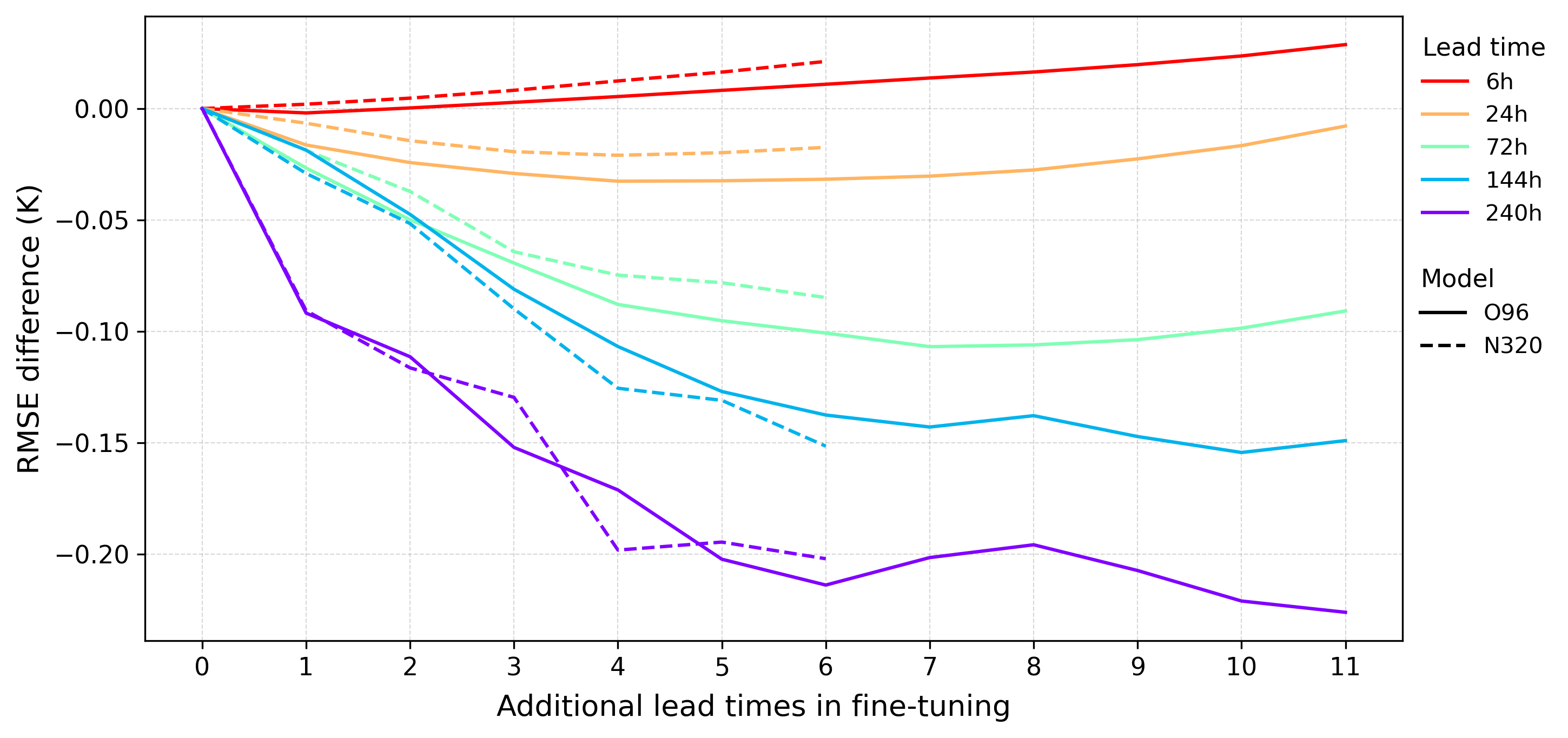}
\caption{The effect of autoregressive fine-tuning on performance at a range of lead times for FastNet O96 and N320. The figure shows RMSE for temperature at 850~hPa for additional lead times in fine-tuning (horizontal axis) at varying forecast lead times (colour). Fine-tuning generally improves RMSE at longer forecast lead times as expected, until seven or eight additional lead times, when it starts increasing at 72~hours and roughly plateaus in the longer lead times. This suggests that the blurring induced by adding additional lead times goes beyond the optimal level and degrades performance. The single step (6~h) RMSE is expected to increase (get worse) with additional fine-tuning, since this is no longer the optimisation target. This is generally observed, except that O96 with one step of fine-tuning shows a slight improvement, suggesting there was more performance that could have been realised during pre-training for this model.}
\label{fig:fine-tuning-t850-difference}
\end{figure}

Figure~\ref{fig:fine-tuning-t850-difference} shows RMSE change for 850~hPa temperature at lead times of 6~h, 24~h, 72~h, 144~h, and 240~h for the O96 and N320 models as a function of additional lead times included in the fine-tuning. As expected, additional lead times in the fine-tuning improve RMSE at longer lead times, at the expense of a slight degradation of performance at 6~h (and then at 24~h for steps beyond 5) as fine-tuning changes the primary training objective. Additional lead times up to seven (equivalent to 48~h lead time) lead to RMSE increasing or plateauing for all forecast lead times, indicating that additional fine-tuning steps do not lead to an overall increase in performance. The N320 model is only fine-tuned out to six additional lead times, and is still showing improved performance as additional steps are added. This is consistent with the observations of fine-tuning on lower resolution models, as Keisler \cite{keisler2022} noted that their $1\degree$ resolution GNN model improves with multi-step rollout loss up to twelve steps, but that the four step model achieved only a slightly worse level of performance when compared to the full twelve step rollout. This may reflect the range over which $1\degree$ and $1.5\degree$ models are expected to see benefit from multi-step fine tuning. The $1.5\degree$ model in ArchesWeather found benefits from fine-tuning even out to four days, though this model has a one day time step and thus the time horizon over which fine-tuning is effective may depend on model time step in addition to the lead time.

\begin{figure}
\includegraphics[width=\textwidth]{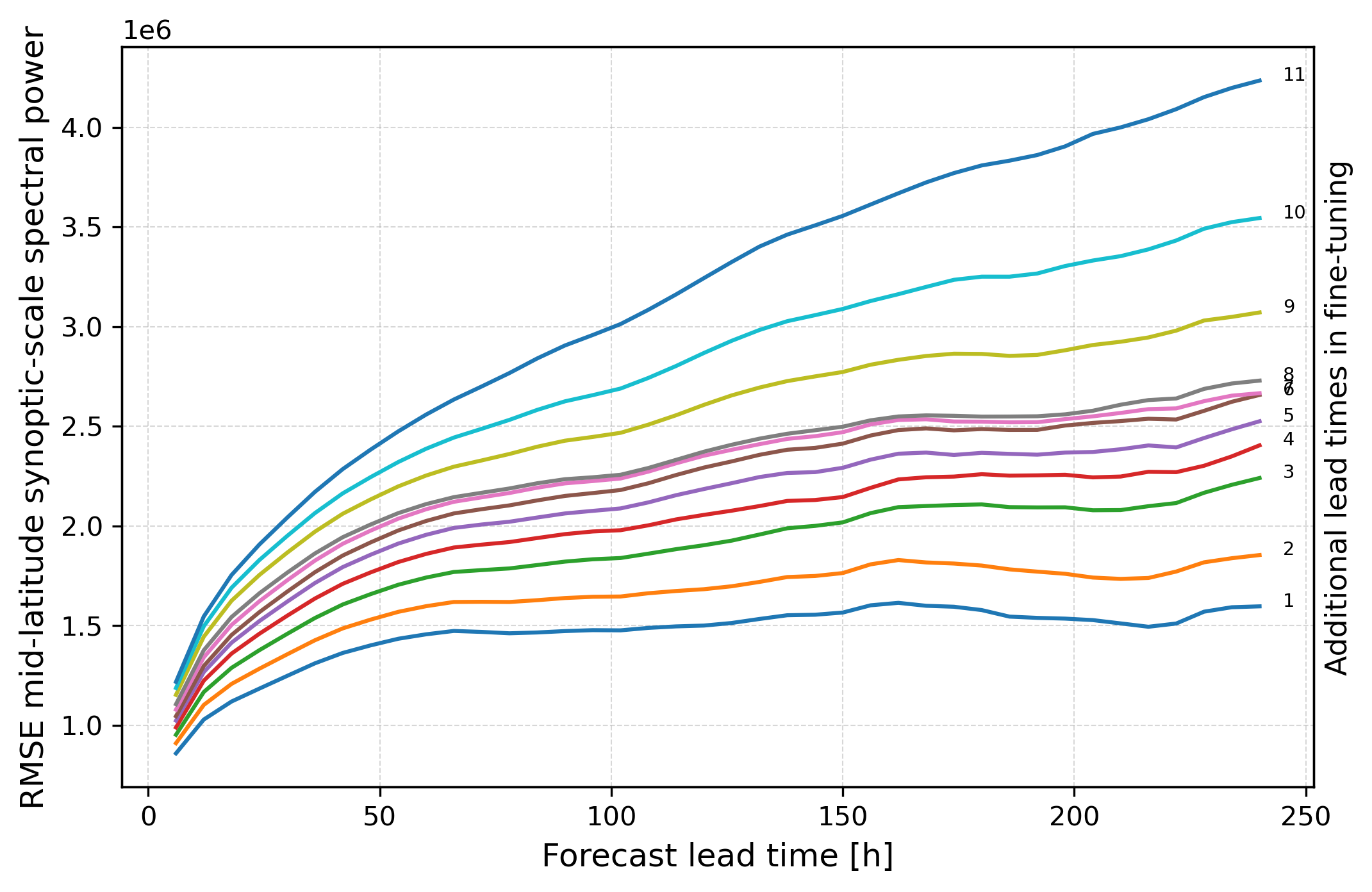}
\caption{RMSE of power spectrum in the synoptic scale (200--2000~km). Lower RMSE indicates a better match to the frequency content in the ground truth data, and a higher value is associated with more blurry spatial features. Notice that this increases with forecast lead time, and tends to plateau. It also increases with additional fine-tuning steps. The lowest RMSE in 850hPa temperature as shown in Fig.~\ref{fig:fine-tuning-t850-difference} occurs after seven or eight additional fine-tuning steps, and appears to `converge' in spectral RMSE.  Additional fine-tuning (9, 10, 11 steps) increases both forecast RMSE and spectral RMSE, suggesting a failure in training.}
\label{fig:fine-tuning-spectral-power}
\end{figure}

As noted in previous work, fine-tuning blurs the forecasts to avoid the well-known double penalty in spatiotemporal prediction problems \cite{rossetal95,ebertetal13,subich_fixing_2025}, which occurs when a high resolution feature is predicted in the wrong place and the model is penalised both for making a high-amplitude feature prediction at an incorrect spatial location and for not predicting the feature in the correct spatial location. Spectral RMSE information is shown for the O96 model in Fig.~\ref{fig:fine-tuning-spectral-power}. The vertical scale shows the RMSE for the mid-latitude synoptic scale spectral power (error computed across a particular frequency band corresponding to a spatial scale ranging from \qtyrange{200}{2000}{km}). Higher values of this spectral RMSE indicate more blurring in the spatial structure of the forecast, while lower values indicate a closer match to the frequency content in the ground truth data. The horizontal scale indicates the forecast lead times, and the lines show behaviour as additional lead times are added in fine-tuning. We see that for the first few lead times, the spectral RMSE in the O96 model grows slowly beyond approximately 48~h lead time, indicating that the forecast reaches a consistent level of blurring over the full 10~day rollout. Adding further lead times steadily increases the frequency-based RMSE up to 8 lead times (54~h), after which the frequency-based RMSE grows much more rapidly. This corresponds to the same number of fine tuning steps as identified in Fig.~\ref{fig:fine-tuning-t850-difference} after which the longer lead time performance plateaus, suggesting that this occurs because the forecast becomes too blurred to provide additional predictive skill when fine-tuned to this number of lead times for the O96 grid.

Based on the above, we conclude that the blurring induced by the multi-step loss is beneficial for the RMSE scores up to 7 auto-regressive steps (6 additional lead times), after which the model predictions are further blurred without producing a corresponding improvement in the mean squared error. In the results that follow, we focus on the O96 model that is fine-tuned up to $T_\mathrm{train}=7$.

\section{Evaluation}

We use the WeatherBench 2 software package \cite{rasp2023weatherbench} to evaluate our forecasts against 2022 as the held out ground truth. We compute the full rollout for all forecast valid times in 2022, re-gridding the FastNet output from the (nominally $1\degree$) O96 grid to a $1.5\degree$ latitude-longitude grid using a conservative re-gridding scheme. Similarly, the outputs from the Met Office Global Model (GM) are also re-gridded to $1.5\degree$ for comparison against the GM's own operational analysis. The GM predictions are initialised at 00:00 UTC and 12:00 UTC each day, with predictions recorded at 12 hour increments up to a 7 day lead time, which are compared to the equivalent predictions from FastNet. Each grid point is area weighted when computing each metric to ensure performance in polar regions does not bias the results.

\begin{figure}[htbp]
  \centering
  \includegraphics[
      width=0.9\linewidth,
      keepaspectratio,
      trim={0cm 0cm 0cm 0cm},  
      clip
  ]{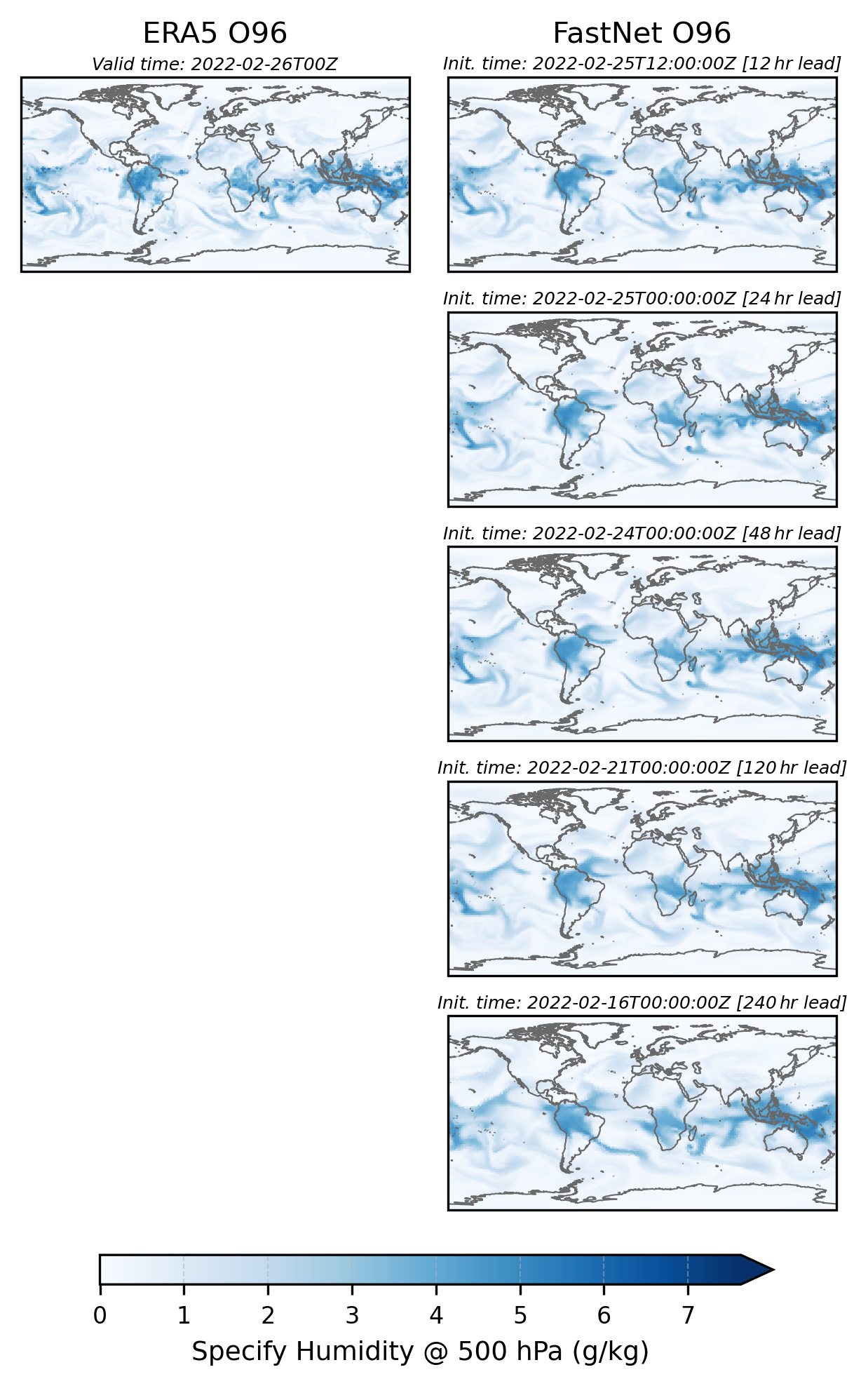}
  \caption{Example predictions of specific humidity at 500~hPa from O96 FastNet for several lead times (6~h, 24~h, 48~h, 120~h, and 240~h, all same valid time) and ground truth ERA5 for comparison. FastNet is able to capture the general weather patterns over the typical lead time range of medium-range weather prediction, with some blurring occurring in the predictions for longer lead times in the fine-tuned model as the model becomes more uncertain.}
  \label{fig:forecast-map}
\end{figure}

An example of the predictions produced by FastNet are shown in Fig.~\ref{fig:forecast-map}. The plots show global maps of specific humidity at 500~hPa for ERA5 and O96 FastNet (fine-tuned) predictions with lead times of 12~h, 24~h, 48~h, 120~h, and 240~h (all at the same valid time). As the lead time increases, the spatial patterns become blurred and features are less well defined as the model become less certain.

The predictions produced by FastNet are then compared against ERA5 data as ground truth and root mean-square-error (RMSE) and anomaly correlation coefficient (ACC) metrics are computed. Evaluation is done using a suite of atmospheric and surface variables as noted in the following, and lead times range from 12 hours up to 168 hours (7 days), the same range of forecasts produced by the GM. We also make predictions out to 10 days with FastNet at 6 hour increments, which are shown in some of the figures that follow.

\section{Results}
\label{sec:results}

\subsection{RMSE Comparisons with Met Office Global Model}

\begin{figure}[htbp]
  \centering
  \includegraphics[
      width=0.9\linewidth,
      keepaspectratio,
      trim={0cm 0cm 0cm 0cm},  
      clip
  ]{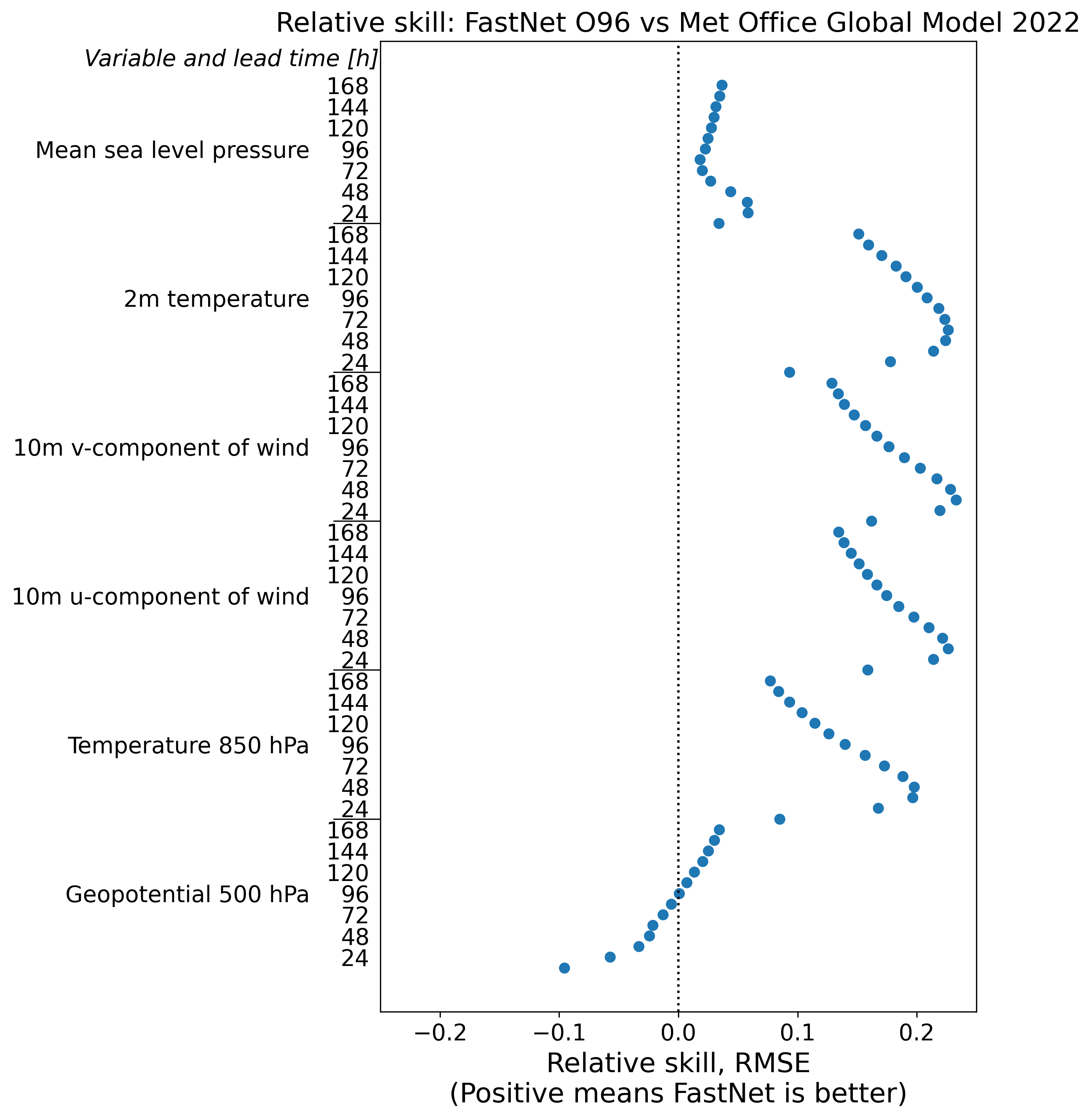}
  \caption{Skill comparison between FastNet (O96) and the Met Office Global Model (GM) for 2022 (forecast valid time). Skill
  is determined by normalising the difference in RMSE between the two models by the GM RMSE for the full year, with positive skill indicating that FastNet has a lower RMSE. FastNet is initialised from and is compared with ERA5 for evaluation, while the GM uses its own operational analysis for both initialisation and ground truth. Comparison variables include geopotential, temperature at 850~hPa, 10~m wind components $u$ and $v$, 2~m temperature, and mean sea level pressure. Lead times range from 12 hours up to 7 days, with a 12 hour increment. For nearly all variables and lead times, FastNet exhibits improved skill over the GM, with the only exception being on geopotential at lead times less than 4 days.}
  \label{fig:scorecard}
\end{figure}

We compare the predictive skill of FastNet (O96 grid) against the Met Office Global Model (GM) measured by RMSE using forecast valid times for the entirety of 2022 in Fig.~\ref{fig:scorecard}. RMSE can be calculated for each level or surface variable according to
\begin{equation}
	{\rm RMSE}_l = \sqrt{\frac{1}{TIJ}\sum_{t}^{T}\sum_{i}^{I}\sum_{j}^{J}w_{i}(x_{tlij}-\hat{x}_{tlij})^2},
\label{eq:rmse}
\end{equation}
where $t,l,i,j$ refer to the time, level, latitude, and longitude indices of the given variable, $x$ is the forecast variable, $\hat{x}$ is the ground truth value, and $w_i$ is the latitude area weight.

\begin{figure*}[!t]
\noindent\includegraphics[width=\textwidth]{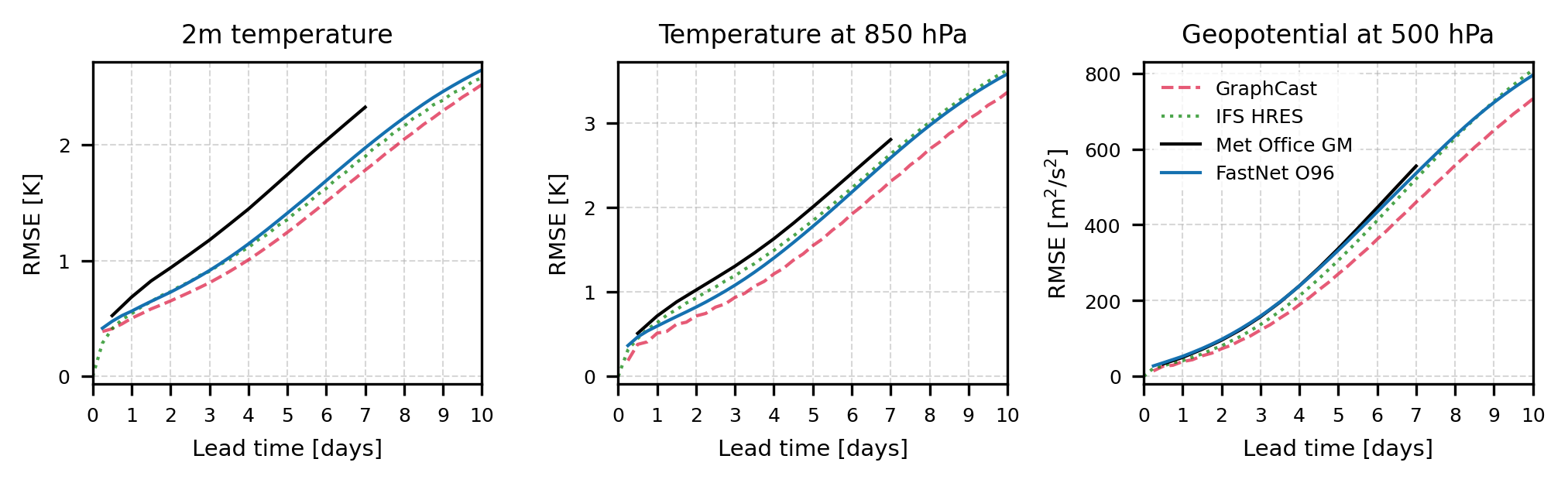}
\caption{RMSE values as a function of lead time for the Met Office Global Model (black) and fine-tuned FastNet O96 model (blue). IFS-HRES and and GraphCast are also shown for comparison. All models are evaluated in 2022, regridded to $1.5\degree$, and compared to the appropriate ground truth dataset.}
\label{fig:RMSE-GM-vs-MSE}
\end{figure*}

Comparisons are made for a range of lead time ranging from 12 hours up to 168 hours (7 days) with a 12 hour increment for a range of atmospheric and surface variables predicted by FastNet. FastNet is initialised using held out ERA5 data from 2022 and evaluated against ERA5, while the GM uses its own operational analysis for initialisation and evaluation. Relative skill is determined by computing RMSE for both models and normalising the difference by the GM RMSE. Positive relative skill indicates that FastNet is performing better with a lower RMSE for the evaluation year. Evaluation variables in this comparison include geopotential at 500~hPa, temperature at 850~hPa, 10~m wind (both $u$ and $v$ components), 2~m temperature, and mean sea level pressure. We also show the RMSE values as a function of lead time for FastNet (both O96 and N320), the GM, IFS-HRES, and GraphCast in Fig.~\ref{fig:RMSE-GM-vs-MSE}. FastNet achieves RMSE values that are competitive with leading physics-based models across a range of metrics and lead times. FastNet does not match the performance of GraphCast, which performs best on the WeatherBench2 benchmark \cite{rasp2023weatherbench}. However, is competitive with the other data-driven models mentioned in Section~\ref{sec:othermodels} which all tend to produce RMSE scores between IFS-HRES and GraphCast \cite{rasp2023weatherbench} across most variables and lead times.

For nearly all variables and lead times, FastNet exhibits superior performance when compared to the GM. The exception is geopotential at 500~hPa, which shows a lower RMSE for lead times up to 96 hours (4 days), where the GM performs better. We find that the largest difference between FastNet and the GM generally occurs around 48 hours lead time, which was the longest lead time over which we perform multi-step fine tuning. This peak in predictive skill matches the final training objective for which the FastNet weights are optimised.

\subsection{Anomaly Correlation Coefficient Comparisons}

\begin{figure}[htbp]
  \centering
  \includegraphics[
      width=0.45\linewidth,
      keepaspectratio,
      trim={0cm 0cm 0cm 0cm},  
      clip
  ]{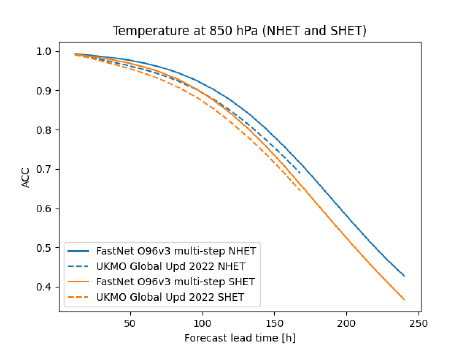}
  \includegraphics[
      width=0.45\linewidth,
      keepaspectratio,
      trim={0cm 0cm 0cm 0cm},  
      clip
  ]{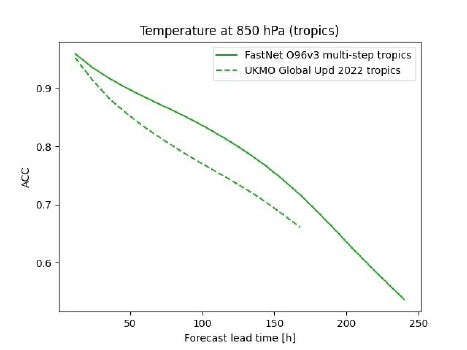}
  \includegraphics[
      width=0.45\linewidth,
      keepaspectratio,
      trim={0cm 0cm 0cm 0cm},  
      clip
  ]{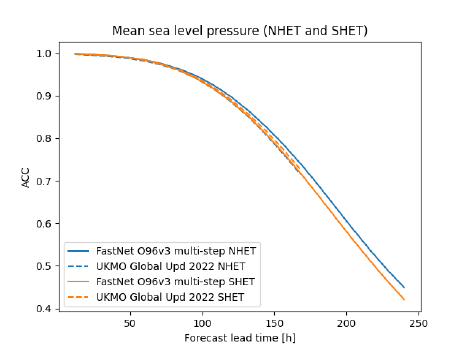}
  \includegraphics[
      width=0.45\linewidth,
      keepaspectratio,
      trim={0cm 0cm 0cm 0cm},  
      clip
  ]{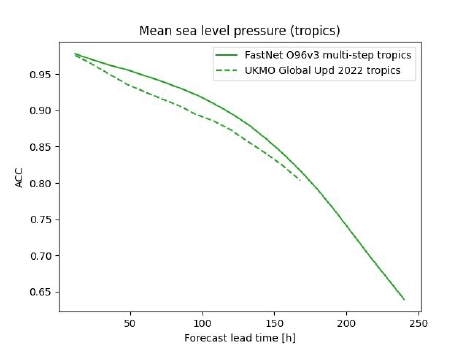}
  \caption{Anomaly Correlation Coefficient scores for FastNet and the GM for 850~hPa temperature and mean sea level pressure (MSLP) computed for Northern/Southern Hemisphere Extra-Tropics (NHET, SHET) and tropics. FastNet shows good performance when measured by ACC and is generally scores better than the GM, with the exception of MSLP SHET.}
  \label{fig:acc}
\end{figure}

To supplement evaluation with RMSE, we also compare both FastNet and the GM using the ACC metric for several sub-regions in Fig.~\ref{fig:acc}:
\begin{equation}
	{\rm ACC}_l=\frac{1}{T}\sum_{t}^{T}\frac{\sum_{i}^{I}\sum_{j}^{J}w_i(x_{tlij}-c_{tlij})(\hat{x}_{tlij}-c_{tlij})}{\sqrt{\sum_{i}^{I}\sum_{j}^{J}w_i(x_{tlij}-c_{tlij})^2}\sqrt{\sum_{i}^{I}\sum_{j}^{J}w_i(\hat{x}_{tlij}-c_{tlij})^2}}.
\end{equation}
where $c_{tlij}$ is the climatology for the given point, where $t$ refers to both day of year and time of day, and $x$ and $\hat{x}$ represent the prediction and ground truth, respectively. ACC measures the correlation between the deviation of the predicted fields and the ground truth from climatology, and ranges from $+1$ (perfectly correlated) to $-1$ (perfectly anti-correlated). A forecast matching climatology has an ACC of 0, and generally if ACC $<0.6$ then the prediction ceases to provide useful positioning of synoptic scale features for prediction purposes.

Comparisons are made for Northern Hemisphere Extra-Tropics (NHET), Southern Hemisphere Extra-Tropics (SHET), and Tropics (defined by latitudes between $30\degree N$ and $30\degree S$S for both 850~hPa temperature and mean sea level pressure (MSLP). With the exception of MLSP for the SHET, the anomaly correlation coefficients for FastNet show an improvement over the GM and generally indicate good spatial correlation between the forecast and ground truth.

\section{Conclusion}                                              

We present the FastNet version 1.0 model, which uses Graph Neural Networks to produce skillful medium range weather forecasts with a global resolution of $1\degree$.  We find that FastNet is able to produce skillful predictions that are competitive with both traditional physics-based NWP models and more recent data-driven models trained on historical data when evaluated retrospectively on a full hold-out evaluation year. While these results are promising, further evaluation of the prospective skill on forecasts using operational analysis is still required before FastNet can be confidently used in operations.

The Met Office is currently publishing a daily experimental 3 day lead time prediction of Mean Sea Level Pressure and 10~m wind vector from FastNet in arrears to continue to monitor the skill of FastNet\footnote{https://www.metoffice.gov.uk/research/approach/collaboration/artificial-intelligence-for-numerical-weather-prediction}. The same prediction is made both with FastNet and the GM, enabling continual comparison with the current Met Office GM as a baseline. Both FastNet and the GM use the GM operational analysis as input in this experimental setting and both models will continue to be updated as further developments are made. This will be used to continue to benchmark the performance of FastNet and look to use it in a full operational setting in the future.

\section*{Acknowledgements}

This work was funded by The Alan Turing Institute and the Met Office. The authors thank Stephen Belcher, Ryan Boult, Jack Bowyer, Maya Bronfeld, Mathew Corbett, Marc Deisenroth, Daniel Delbarre, Anna-Louise Ellis, Ben Fitzpatrick, Mark Girolami , Nicolas Guernion, Stephen Haddad, Richard Hattersley, Tom Henderson, Aaron Hopkinson, Richard Lawrence, Tomas Lazauskas, Clodagh Lynch, Ben MacArthur, Cyril Morcrette, Kelly O'Meara, Martin O'Reilly, Joseph Palmer, James Penn, Bernat Puig-Camps, Christine Sheldon, Jonathan Starck, Hannah Sweeney, Richard Turner, Monica Vakil-Dewar, Luke Vinton, Simon Vosper, George Williams, and Keith Williams for their support in this work.

\bibliographystyle{plain}
\bibliography{bib}

\appendix

\section{Graph Neural Networks}\label{appendix:gnns}

A Graph Neural Network (GNN) is a graph structure that contains features---on any of the nodes, edges, and graph itself---that can be updated by learnable functions. These functions are first applied and shared across edges, then applied, shared, and aggregated across node neighbourhoods, This procedure is known as \emph{message passing} to indicate that information is propagating throughout the graph.

All GNNs in this paper are \emph{interaction networks} \cite{interaction-net}, which have been successfully used to model the dynamics of physical processes. We have defined the specific graphs used in preceding sections of the main text (encoder, mesh, decoder), but they all follow the same update procedure that we detail here.

Given a feature vector $\mathbf{x_u}$ for node $u$, and an edge feature vector $\mathbf{x_{uv}}$ between nodes $u$ and $v$, an interaction network will first update the edge features:
\begin{equation*}
    \mathbf{h_{uv}} = \psi \left( \text{concat} \left( \mathbf{x_u}, \mathbf{x_v}, \mathbf{x_{uv}}\right) \right) ~\text{for all edges }\mathbf{uv},
\end{equation*}
where $\psi$ is a single layer MLP shared over all edges, and $\text{concat}(\cdot)$ is the concatenation operation along the feature axis, which combines the different feature vectors into a single larger vector for the MLP.

These updated edge messages are aggregated at the original node, and concatenated with the original node features $\mathbf{x_u}$ before being updated:
\begin{equation*}
   \mathbf{h_u} = \phi \left( \text{concat} \left(  \mathbf{x_u}, \bigoplus_{v \in \mathcal{N}_u} \mathbf{h_{vu}} \right) \right), 
\end{equation*}
where $\phi$ is a 1-layer MLP shared over all nodes, $\mathcal{N}_u$ is the 1-hop neighborhood of node $u$, and $\bigoplus$ represents a general aggregation operator, which we implement in practice as the sum. Note that if the graph is directed, the neighborhood of $u$ will only contain edges pointing directly to $u$, and not those that point away from $u$.

We can imagine these latent variables $\mathbf{h_{uv}}$ and $\mathbf{h_{u}}$ to then replace their original counterparts in-place, i.e. $\mathbf{x_{uv}} \leftarrow \mathbf{h_{uv}}$ and $\mathbf{x_{u}} \leftarrow \mathbf{h_{u}}$, which allows us to repeatedly apply this formulation. Composing “layers” of these operations allows the GNN to be more expressive, and is done by introducing new MLPs for each layer, which can differ in output sizes, and will have different sets of learnable weights — this would introduce subscripts $\psi_i$, $\phi_i$ for layer $i$. This process is also referred to as having multiple rounds of message passing. We also use residual connections between each round, where the original input $\mathbf{x_{u}}$  is added to the result $\mathbf{h_{u}}$.

\section{Encoder Graph Structure}\label{appendix:encoder}

\begin{figure*}[t]
\noindent\includegraphics[width=\textwidth]{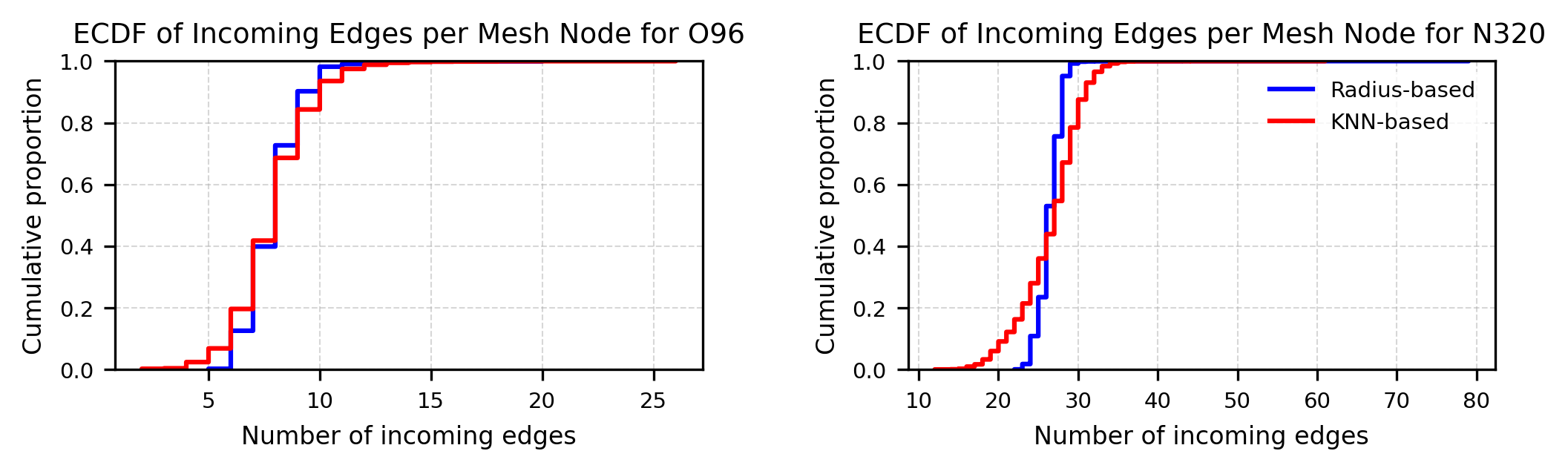}
\caption{Radius-based vs nearest neighbour encoders. Panels (A) and (B) show the cumulative distribution functions (ECDFs) of incoming node degrees on the mesh for the O96 and N320 configurations, respectively, comparing the radius-based (blue) and K-nearest-neighbours (KNN; red) encoder graph construction methods with $k=2$. All FastNet models use a KNN structure for the decoder graph, as is also done in other GNN-based data-driven weather models \cite{Lam2023,Lang2024}.}
\label{fig:ECDF-incoming-edges-encoder}
\end{figure*}

As noted in the main text, we evaluate KNN and radius-based connectivity approaches in the encoder graph for both the O96 and N320 models. For KNN models, grid points are connected to their $k$-nearest mesh points, with $k=2$ used for the encoder graph. Radius models use a fixed distance of \qty{89}{km} or \qty{178}{km} for the N320 and O96 models, respectively, or approximately 0.8 times the mesh resolution. Although KNN yields consistent connectivity, with a constant number of outgoing edges from each grid node, it tends to produce a wider variation of number of incoming edges for each mesh node. We also implement the radius-based strategy to evaluate its impact and compared it against KNN for the O96 and N320 models to see whether the edge-construction strategy affects prediction accuracy. More precisely, the number of edges from each grid point to the mesh depends on the selected mechanism---either a fixed $k$ or a geodesic radius. Figure~\ref{fig:ECDF-incoming-edges-encoder} shows the cumulative distribution function of number of incoming graph edges connected to each mesh node for the (a) O96 and (b) N320 models. The distribution of incoming edges is sharper under the radius-based method than under KNN, which is especially evident in the N320 model.

This results in a slightly different graph structure when compared to a radius-based approach, where each mesh point is connected to all grid points within a certain radius (usually chosen to be slightly less than the resolution of the mesh, in this case a factor of 0.8 smaller than the finest mesh resolution). These differences are minor for the O96 grid ($\theta\approx2\degree$ mesh resolution), and more pronounced when using the N320 grid and $\theta\approx1\degree$ as illustrated in Fig.~\ref{fig:ECDF-incoming-edges-encoder}. We find that both approaches produce similar results in the predictive skill of the final model on the O96 grid, while the radius method performs slightly better on the N320 grid. As noted in the main text, these differences are of approximately 2-10\% of the KNN RMSE score for most variables, with smaller relative differences at long lead times and larger relative differences at short lead times.

\clearpage
\includepdf{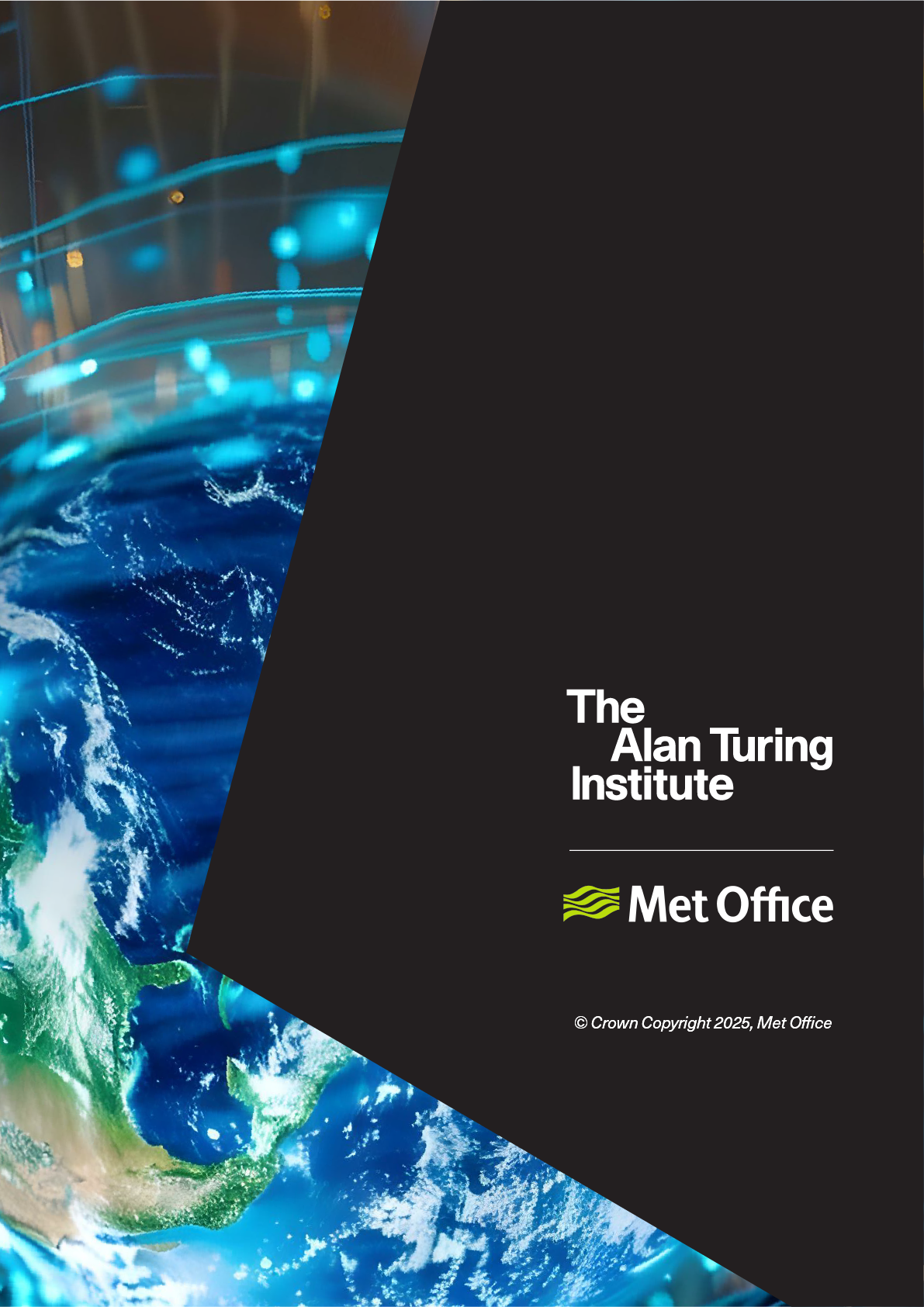}

\end{document}